\documentclass[12pt]{article}
\usepackage{amsmath}
\usepackage{graphicx}
\usepackage{cite}
\usepackage{hyperref}
\usepackage{geometry}
\title{Revolutionizing Pharmaceutical Manufacturing: Advances and Challenges of 3D Printing System and Control}
\author{Rahul Kumar, Vikram Singh and Priya Gupta \\
Department of Biotechnology, Bharathiar University, India \\
\texttt{priya.gupta103@b-u.ac.in}}
\date{}

\begin{document}

\maketitle

\begin{abstract}
    The advent of 3D printing has transformed the pharmaceutical industry, enabling precision drug manufacturing with controlled release profiles, dosing, and structural complexity. Additive manufacturing (AM) addresses the growing demand for personalized medicine, overcoming limitations of traditional methods. This technology facilitates tailored dosage forms, complex geometries, and real-time quality control. Recent advancements in drop-on-demand printing, UV curable inks, material science, and regulatory frameworks are discussed. Despite opportunities for cost reduction, flexibility, and decentralized manufacturing, challenges persist in scalability, reproducibility, and regulatory adaptation. This review provides an in-depth analysis of the current state of AM in pharmaceutical manufacturing, exploring recent developments, challenges, and future directions for mainstream integration.
\end{abstract}

\section{Introduction}
The advent of 3D printing has revolutionized the pharmaceutical industry, offering unprecedented precision, customization, and efficiency in drug manufacturing. Additive manufacturing (AM), also known as 3D printing, has emerged as a game-changer, enabling layer-by-layer fabrication of drug products with controlled release profiles, dosing, and structural complexity. This transformative technology has the potential to address the growing demand for personalized medicine, where traditional manufacturing methods often fall short.

Traditional manufacturing methods excel in large-scale production but often lack flexibility for complex, patient-specific dosage forms. In contrast, AM offers unparalleled control over drug release profiles, dosing, and structural complexity, enabling the creation of tailored dosage forms that meet specific therapeutic needs. Moreover, AM facilitates the production of complex geometries, such as multi-compartmental systems, and enables the incorporation of multiple active pharmaceutical ingredients (APIs) within a single dosage form.

Improving production processes involves not only enhancing efficiency but also ensuring rigorous control over product quality and consistency. Techniques like drop-on-demand (DoD) printing and UV curable inks, originally developed for industrial applications, are being adapted for pharmaceutical manufacturing. These methods enable precise deposition of drug-laden materials, allowing for tailored dosage forms that meet specific therapeutic needs. Furthermore, AM enables real-time monitoring and adjustment of production parameters, ensuring consistent product quality.

Recent research has focused on refining the accuracy and reliability of AM processes. The quality of 3D-printed drug products depends on controlling the height profile of each printed layer, which affects drug release uniformity and reproducibility. Improved models for predicting height profiles and new control strategies, such as feedforward and feedback mechanisms, have been developed to mitigate errors and maintain consistency. Additionally, advancements in material science have led to the development of novel pharmaceutical-grade materials suitable for 3D printing.

The integration of AM in pharmaceutical manufacturing also offers opportunities for cost reduction and increased flexibility. By eliminating the need for molds and tooling, AM enables rapid prototyping and production of small batches, reducing development costs and lead times. Moreover, AM facilitates decentralized manufacturing, enabling production of medicines at or near the point of care.

However, challenges remain in translating AM technology into mainstream pharmaceutical manufacturing. Regulatory frameworks must be adapted to accommodate the unique characteristics of 3D-printed medicines, ensuring quality, safety, and efficacy. Scalability and reproducibility must also be demonstrated to meet commercial demands.

This review provides an in-depth analysis of the current state of drug manufacturing using AM techniques, focusing on DoD printing and UV curable inks. It discusses recent developments in:
\begin{enumerate}
\item Height profile modeling and control strategies
\item Pharmaceutical-grade material development
\item Regulatory frameworks and quality control
\item Scalability and reproducibility
\end{enumerate}

The review explores challenges and future directions for integrating AM technologies into mainstream pharmaceutical manufacturing, emphasizing regulatory compliance, operational scalability, and patient-centric innovation.

\section{Additive Manufacturing Methods and Their Role in Drug Manufacturing}
Additive manufacturing (AM), commonly known as 3D printing, encompasses a variety of techniques that build objects layer by layer based on digital models. Initially developed for industrial applications, AM has found significant relevance in healthcare and pharmaceuticals, especially for the production of complex, customizable drug delivery systems. The flexibility and precision of these technologies offer new possibilities for manufacturing personalized and patient-specific dosage forms. This section highlights the key AM techniques used in drug manufacturing and their potential impact on pharmaceutical production.

\subsection{Fused Deposition Modeling (FDM)}
Fused deposition modeling (FDM) is a well-established 3D printing technique that uses thermoplastic filaments extruded through a heated nozzle to create objects layer by layer. In pharmaceutical applications, FDM is primarily used to fabricate solid oral dosage forms, including tablets and capsules, which can be customized to achieve precise control over drug release profiles. The ability to produce multi-drug dosage forms with specific geometries enables the creation of patient-specific medications, addressing personalized treatment needs \cite{sathies2020review,jamroz20183d,wu2016command,park2019pharmaceutical}. However, the high temperatures required for filament extrusion may degrade heat-sensitive active pharmaceutical ingredients (APIs), making careful selection and optimization of excipients critical \cite{parulski2021challenges}. Research has also explored the use of polymer blends and drug-polymer mixtures to enhance drug stability and bioavailability in FDM-printed products \cite{wang20223d}.

\subsection{Stereolithography (SLA)}
Stereolithography (SLA) utilizes a laser to cure liquid photopolymer resin layer by layer, enabling the production of high-resolution structures with complex geometries. In pharmaceuticals, SLA has been employed to fabricate drug delivery devices, implants, and other dosage forms that require intricate internal structures for controlled drug release. Recent advancements in photopolymer materials have expanded the range of APIs that can be incorporated, though concerns remain regarding the biocompatibility and safety of these materials \cite{ge2020projection,noh20183d}. SLA’s precision and ability to produce devices with complex, internal channels make it ideal for applications where spatial distribution of the drug is crucial. However, post-processing steps, such as cleaning and additional curing, add complexity and time to the overall manufacturing process \cite{chhikara2021bioactive}.

\subsection{Selective Laser Sintering (SLS)}
Selective laser sintering (SLS) involves the use of a laser to sinter powdered materials, layer by layer, creating solid structures. The absence of solvents and binders in SLS makes it suitable for APIs that are sensitive to moisture, solvents, or chemical reactions. Moreover, SLS is capable of producing porous structures, which are advantageous for controlled drug release applications \cite{fina2017selective}. The ability to fine-tune porosity and density within the printed object allows for the development of dosage forms with customizable release profiles, making SLS an attractive option for extended-release formulations. Despite its potential, the high temperatures involved in the sintering process can pose challenges for thermally labile drugs, requiring careful control over process parameters \cite{vock2019powders}.

\subsection{Inkjet Printing}
Inkjet printing, particularly drop-on-demand (DoD) printing, is a highly versatile technique that precisely deposits droplets of drug-containing inks onto a substrate. Unlike other 3D printing methods, inkjet printing operates at ambient or low temperatures, making it particularly suitable for heat-sensitive APIs. This low-temperature process preserves the stability and bioactivity of sensitive drugs, such as biologics and peptides \cite{wu2019modeling,carou2024inkjet,daly2015inkjet}. 

Moreover, inkjet printing offers exceptional control over dosage, spatial distribution, and layering, enabling the production of multi-drug formulations and complex drug release profiles in a single dosage form \cite{goyanes2015personalized}. The use of non-toxic solvents, such as water or ethanol, which evaporate during the curing process, ensures the safety and biocompatibility of the final product \cite{bernard2018biocompatibility}. Recent advancements include the development of UV-curable inks that enhance the mechanical strength and stability of printed formulations while maintaining high precision \cite{wu2021improved,carou2024inkjet}. Inkjet printing’s versatility extends to the production of orodispersible films, microneedles, and other drug delivery systems that require precise control over dose and structure \cite{chou2021inkjet}.

Overall, inkjet printing’s low temperature processing, non-toxic solvents, and precision make it one of the most promising additive manufacturing techniques for pharmaceutical applications, particularly in the context of personalized and on-demand drug production.

\subsection{Powder Bed Fusion (PBF)}
Powder bed fusion (PBF) techniques, such as multi-jet fusion and selective laser sintering (SLS), involve spreading a layer of powder and selectively fusing it to create solid objects. PBF techniques are particularly effective for producing porous drug delivery systems that can modulate drug release based on the internal structure’s geometry and porosity \cite{vock2019powders,wang2023review}. The high degree of control over particle fusion and distribution enables the development of complex dosage forms with programmed release profiles, making PBF a promising approach for extended-release and targeted drug delivery applications \cite{borandeh2021polymeric}. Additionally, the scalability of PBF techniques offers potential for decentralized and on-demand drug manufacturing, addressing logistical challenges in remote areas or during emergency scenarios \cite{seoane2023case,lupi2023blockchain}.

\subsection{Hot Melt Extrusion (HME)}
Hot melt extrusion (HME) is commonly integrated with fused deposition modeling (FDM) in pharmaceutical applications. In HME, drug and polymer mixtures are processed at high temperatures and pressures to produce a homogeneous extrudate, which can then be shaped into dosage forms. HME is particularly effective for enhancing the solubility and bioavailability of poorly water-soluble APIs by converting them into amorphous solid dispersions \cite{melocchi2016hot,park2019pharmaceutical}. When combined with 3D printing, HME allows for the production of complex dosage forms with tailored drug release profiles. However, like FDM, the elevated temperatures used in HME can limit its applicability to heat-stable APIs \cite{wang2021emerging}.

\subsection{Relationship with Drug Manufacturing}
Additive manufacturing technologies have the potential to revolutionize pharmaceutical production by enabling the development of personalized, complex, and efficient drug delivery systems. The key benefits include:
\begin{itemize}
    \item 
AM technologies provide unparalleled flexibility in creating dosage forms tailored to individual patient needs. This is particularly valuable in personalized medicine, where therapies are customized based on patient-specific factors such as age, weight, and genetic profile \cite{govender2020independent,seoane2023case, wu2021improved2}. For instance, inkjet printing’s precise dosing capability allows for the creation of formulations with highly individualized doses, improving therapeutic outcomes \cite{goyanes2015personalized}.

    \item AM allows for the design and manufacture of drug delivery systems with intricate internal architectures and programmable release profiles. Techniques such as SLA and inkjet printing excel in producing multi-drug formulations, implants, and scaffolds with precise control over spatial drug distribution and release kinetics \cite{melocchi20153d,trivedi2018additive}. The ability to integrate multiple drugs within a single dosage form and control their release offers significant advantages for complex treatment regimens.

    \item The flexibility and scalability of AM make it suitable for decentralized, on-demand drug production. This is particularly beneficial in remote or resource-limited settings, where access to centralized drug manufacturing facilities may be restricted \cite{seoane2023case,lupi2023blockchain}. AM also offers potential for reducing waste and optimizing supply chains by producing only the quantities needed, thereby improving overall manufacturing efficiency \cite{attaran2017additive}.

    \item Despite its promise, the adoption of AM in drug manufacturing faces significant regulatory and scalability hurdles. Ensuring consistent quality, process validation, and compliance with stringent regulatory requirements are critical issues that must be addressed for broader industry adoption \cite{rafi2022regulatory}. Collaborative efforts between regulatory bodies, academic institutions, and industry stakeholders are necessary to establish standardized guidelines and validation protocols for AM-based pharmaceutical production.

\end{itemize}
In conclusion, the versatility and precision of additive manufacturing methods are driving innovation in drug manufacturing, offering opportunities to develop advanced, personalized therapies. However, successful integration into commercial pharmaceutical production will require continued advancements in process control, material science, and regulatory frameworks.

\section{Challenges and Latest Developments in Height and Geometry Profile Modeling for Inkjet Printing in Drug Manufacturing}

Inkjet printing has emerged as a versatile and promising technique in drug manufacturing due to its ability to precisely deposit small volumes of drug-containing inks onto substrates. However, controlling the height and geometry of the printed structures remains a significant challenge, particularly when aiming to produce dosage forms with complex shapes, accurate dosages, and consistent release profiles. The following section discusses the main challenges associated with height and geometry profile modeling in inkjet printing, along with recent advances in addressing these issues.

\subsection{Droplet Formation and Deposition Dynamics}
One of the fundamental challenges in inkjet printing is achieving consistent droplet formation and deposition. The shape and size of the deposited droplet are influenced by factors such as ink viscosity, surface tension, and printhead dynamics. Variations in these parameters can lead to inconsistent droplet sizes, resulting in height variations and non-uniform layer deposition, which are critical for controlling dosage and release profiles in drug manufacturing \cite{geraili2021design,wu2022height,cailleaux2021fused,kapoor2020coating}.

Recent developments in drop-on-demand (DoD) inkjet printing have focused on improving the control over droplet formation through optimized printhead designs and ink formulations. Advanced models now account for the dynamic interplay between fluid properties and printhead actuation, providing better predictions of droplet behavior and height profiles \cite{wu2021improved2}. Additionally, techniques such as multi-layer deposition and sequential curing have been employed to reduce height inconsistencies and achieve smoother surface geometries, which are essential for ensuring uniform drug distribution in printed dosage forms \cite{wu2022height}.

\subsection{Surface Interactions and Substrate Properties}
The interaction between the deposited droplet and the substrate plays a crucial role in determining the final geometry and height of the printed structure. Factors such as wetting behavior, spreading, and absorption influence the droplet's shape and can lead to unintended variations in the layer thickness. For drug manufacturing, where precise control over dosage is critical, any deviation in the geometry of the printed layer can compromise the therapeutic efficacy and safety of the final product.

To address these challenges, researchers have developed advanced models that incorporate surface energy considerations and substrate properties into height and geometry predictions. Recent studies have introduced improved wetting models that simulate the spreading and absorption of droplets on various substrates, enabling more accurate predictions of the final printed structure \cite{wu2019modeling}. Moreover, the development of UV-curable inks, which solidify rapidly upon exposure to light, has provided greater control over the geometry of printed layers, minimizing the impact of surface interactions and enabling more consistent height profiles \cite{carou2024inkjet}.

\subsection{Curing and Solidification Dynamics}
The curing process, particularly in UV-curable ink systems, introduces additional complexities in height and geometry control. As the ink solidifies, factors such as polymerization shrinkage and curing gradients can lead to distortions in the printed structure. In drug manufacturing, these distortions can affect the uniformity of drug loading and release kinetics, making it essential to develop predictive models that account for these curing-related effects \cite{wu2023error}.

Recent advances have focused on integrating curing dynamics into height profile models. By simulating the photopolymerization process, researchers can better predict how factors like light intensity, exposure time, and ink composition influence the final geometry. Error diffusion-based feedforward control strategies have also been developed to dynamically adjust printing parameters in real-time, compensating for deviations caused by curing and solidification effects \cite{wu2023error}. These approaches are particularly useful for achieving the high precision required in drug formulations, where even minor variations in geometry can impact drug release behavior.

\subsection{Multi-Layer and Complex Geometry Fabrication}
Producing multi-layer structures with consistent height and geometry is a key requirement for drug delivery systems that require complex release profiles. However, layer stacking introduces challenges such as cumulative height errors, inter-layer adhesion issues, and non-linear distortions, which can affect the overall performance of the dosage form. In the context of drug manufacturing, where each layer may contain different drug components or concentrations, maintaining consistent geometry across multiple layers is critical for ensuring therapeutic accuracy \cite{wu2022height}.

To tackle these challenges, researchers have explored improved stacking algorithms and height correction models that compensate for layer-specific variations. Additionally, novel approaches such as adaptive layer thickness control and variable droplet size printing have been developed to enhance multi-layer fabrication. These techniques allow for the fine-tuning of layer heights and enable the construction of complex geometries with precise control over each layer’s contribution to the overall structure.

\subsection{Process Control and Real-Time Monitoring}
Real-time monitoring and adaptive control of the printing process are essential for ensuring the consistency and reliability of printed drug products. Variations in environmental conditions, such as temperature and humidity, as well as ink aging, can impact the droplet formation, deposition, and curing processes. Developing robust process control strategies that can adapt to these variations is crucial for achieving the desired height and geometry profiles in drug manufacturing applications \cite{wu2023error}.

Recent advances in this area include the integration of machine learning algorithms and optical feedback systems for real-time process monitoring. These systems can detect deviations from the expected height profiles and adjust printing parameters on the fly, reducing the likelihood of defects and ensuring consistent product quality. Such advancements are particularly important for large-scale manufacturing, where process variability needs to be minimized to meet regulatory requirements and ensure patient safety \cite{boehm2014inkjet}.

Conclusion
The challenges associated with height and geometry profile modeling in inkjet printing are central to the successful application of this technology in drug manufacturing. Recent developments in droplet dynamics, surface interaction modeling, curing processes, multi-layer fabrication, and process control have significantly improved the accuracy and reliability of printed dosage forms. As these models continue to evolve, the pharmaceutical industry is likely to see broader adoption of inkjet printing for the production of personalized medicines, complex drug delivery systems, and other advanced therapeutic products.

\section{Discussion and Future Directions}
Despite significant advancements in height and geometry profile modeling for inkjet printing, several challenges remain that hinder the widespread adoption of this technology in drug manufacturing. Future research efforts are expected to focus on addressing these challenges and pushing the boundaries of what can be achieved with inkjet printing in pharmaceutical applications.

\subsection{Integration of Advanced AI and Machine Learning Models}
The complexity of accurately predicting height and geometry profiles in inkjet printing requires sophisticated computational models that can handle the dynamic interactions between ink properties, substrate characteristics, and environmental conditions. Incorporating machine learning (ML) and artificial intelligence (AI) into existing modeling frameworks could enable more accurate and adaptive control over the printing process. By analyzing large datasets generated from multiple print trials, ML algorithms can identify patterns and predict outcomes more precisely, leading to better geometry control and reduced variability in printed drug products. Future work could also explore the development of closed-loop control systems powered by real-time AI feedback, which would continuously optimize printing parameters during production \cite{wu2023error,boehm2014inkjet}.

\subsection{Multi-Material and Complex Drug Delivery Structures}
Current research primarily focuses on single-material inkjet printing; however, many pharmaceutical applications require multi-material systems, especially for advanced drug delivery devices such as multi-layer tablets, controlled-release systems, or combination therapies. Future work will likely emphasize the modeling and process control needed for printing with multiple inks, each with distinct rheological and curing characteristics. Addressing the interactions between different materials and how they affect the final height and geometry will be a critical research area. Additionally, the ability to print intricate, high-resolution geometries for specialized drug release profiles, such as pulsatile or delayed-release systems, will demand more sophisticated height profile models that account for complex layering and material blending \cite{parulski2021challenges,goyanes2015personalized}.

\subsection{Advanced Curing and Post-Processing Techniques}
UV-curable inks have already improved height and geometry control in inkjet printing, but further advancements are needed in the curing process to enhance the resolution and mechanical properties of printed drug products. Future research could investigate novel photopolymerization strategies, including dual-curing systems or sequential curing processes, to achieve better layer adhesion and minimize defects like warping and shrinkage. Another promising area is the development of hybrid printing systems that combine inkjet printing with other additive manufacturing methods, such as stereolithography or hot melt extrusion, to create multi-functional dosage forms with intricate geometries and enhanced performance characteristics \cite{ge2020projection,noh20183d}.

\subsection{Personalized Medicine and On-Demand Manufacturing}
One of the most exciting applications of inkjet printing is its potential for personalized medicine, where customized drug formulations and dosages can be printed on-demand based on individual patient needs. As this field evolves, future work will need to focus on developing robust, scalable manufacturing systems that can produce highly personalized drug products while maintaining strict regulatory compliance. Key challenges include ensuring the reproducibility and stability of printed formulations over time, as well as developing user-friendly interfaces for healthcare providers to quickly design and manufacture patient-specific medications \cite{seoane2023case,govender2020independent}. In addition, decentralized manufacturing in remote or resource-limited settings could become more feasible with the integration of automated, AI-driven inkjet systems capable of producing a wide range of dosage forms.

\subsection{Standardization and Regulatory Pathways}
As inkjet printing moves from the lab to the manufacturing floor, there will be an increasing need for standardized protocols and regulatory frameworks that ensure product quality and safety. Regulatory bodies are already exploring guidelines for 3D-printed pharmaceuticals, but the unique challenges of inkjet printing—such as variability in layer deposition and the use of novel materials—will require tailored approaches. Future work should focus on developing standardized testing methods for assessing the mechanical properties, dissolution rates, and stability of inkjet-printed drug products. Additionally, establishing robust quality control procedures that leverage in-line monitoring and real-time data analysis will be essential for regulatory approval and large-scale commercialization \cite{rafi2022regulatory}.

Conclusion
The future of inkjet printing in drug manufacturing is promising, with ongoing research expected to bring about transformative advancements in personalized medicine, multi-material printing, and AI-driven process control. Overcoming the challenges associated with height and geometry profile modeling will be critical for unlocking the full potential of this technology, paving the way for the production of highly customized, complex, and safe pharmaceutical products. Continued collaboration between material scientists, engineers, and pharmaceutical experts will be key to driving innovation and establishing inkjet printing as a cornerstone of next-generation drug manufacturing.

\section{Conclusion}

Inkjet printing is rapidly gaining recognition as a viable and versatile technology for drug manufacturing, offering unprecedented control over dosage forms and enabling personalized medicine. By leveraging the precision of droplet-based deposition, this method allows for the fabrication of complex geometries, multi-layered structures, and customized drug release profiles. However, ensuring consistent quality in printed dosage forms remains a challenge, particularly when it comes to controlling the height and geometry of printed structures.

Significant progress has been made in height and geometry profile modeling, with advancements in droplet dynamics, surface interaction modeling, curing processes, and process control. Recent developments, such as machine learning-driven models, real-time feedback systems, and multi-material printing approaches, are improving the accuracy and reliability of printed drug products. As researchers continue to refine these models, the future of inkjet printing in pharmaceuticals is expected to bring even more sophisticated solutions, making it possible to produce highly customized, on-demand medications that are both safe and effective.

Looking ahead, the integration of AI, improved multi-material systems, and novel curing techniques are likely to push the boundaries of what can be achieved with inkjet printing. Additionally, the regulatory landscape will need to evolve alongside these technological advancements to ensure the safe and efficient deployment of this manufacturing method in clinical and commercial settings. Continued research and collaboration across disciplines will be key to addressing these challenges and unlocking the full potential of inkjet printing in drug manufacturing, ultimately transforming the landscape of pharmaceutical production.

In summary, inkjet printing represents a powerful tool in the pharmaceutical industry, offering tailored drug products with precise control over dosage and release profiles. As the technology advances, it is poised to become a cornerstone of personalized medicine and decentralized manufacturing, delivering significant benefits to both patients and healthcare providers.

\bibliographystyle{ieeetr}
\bibliography{references}

\begin{thebibliography}{10}

\bibitem{sathies2020review}
T.~Sathies, P.~Senthil, and M.~Anoop, ``A review on advancements in
  applications of fused deposition modelling process,'' {\em Rapid Prototyping
  Journal}, vol.~26, no.~4, pp.~669--687, 2020.

\bibitem{jamroz20183d}
W.~Jamr{\'o}z, J.~Szafraniec, M.~Kurek, and R.~Jachowicz, ``3d printing in
  pharmaceutical and medical applications--recent achievements and
  challenges,'' {\em Pharmaceutical research}, vol.~35, pp.~1--22, 2018.

\bibitem{wu2016command}
Y.~Wu, ``Command shaping with constrained peak input acceleration to minimize
  residual vibration in a flexible-joint robot,'' Master's thesis, Purdue
  University, 2016.

\bibitem{park2019pharmaceutical}
B.~J. Park, H.~J. Choi, S.~J. Moon, S.~J. Kim, R.~Bajracharya, J.~Y. Min, and
  H.-K. Han, ``Pharmaceutical applications of 3d printing technology: current
  understanding and future perspectives,'' {\em Journal of Pharmaceutical
  Investigation}, vol.~49, pp.~575--585, 2019.

\bibitem{parulski2021challenges}
C.~Parulski, O.~Jennotte, A.~Lechanteur, and B.~Evrard, ``Challenges of fused
  deposition modeling 3d printing in pharmaceutical applications: where are we
  now?,'' {\em Advanced drug delivery reviews}, vol.~175, p.~113810, 2021.

\bibitem{wang20223d}
N.~Wang, H.~Shi, and S.~Yang, ``3d printed oral solid dosage form: Modified
  release and improved solubility,'' {\em Journal of Controlled Release},
  vol.~351, pp.~407--431, 2022.

\bibitem{ge2020projection}
Q.~Ge, Z.~Li, Z.~Wang, K.~Kowsari, W.~Zhang, X.~He, J.~Zhou, and N.~X. Fang,
  ``Projection micro stereolithography based 3d printing and its
  applications,'' {\em International Journal of Extreme Manufacturing}, vol.~2,
  no.~2, p.~022004, 2020.

\bibitem{noh20183d}
S.~Noh, N.~Myung, M.~Park, S.~Kim, S.-U. Zhang, and H.-W. Kang, ``3d
  bioprinting for tissue engineering,'' {\em Clinical Regenerative Medicine in
  Urology}, pp.~105--123, 2018.

\bibitem{chhikara2021bioactive}
N.~Chhikara, A.~Kaur, S.~Mann, M.~Garg, S.~A. Sofi, and A.~Panghal, ``Bioactive
  compounds, associated health benefits and safety considerations of moringa
  oleifera l.: An updated review,'' {\em Nutrition \& Food Science}, vol.~51,
  no.~2, pp.~255--277, 2021.

\bibitem{fina2017selective}
F.~Fina, A.~Goyanes, S.~Gaisford, and A.~W. Basit, ``Selective laser sintering
  (sls) 3d printing of medicines,'' {\em International journal of
  pharmaceutics}, vol.~529, no.~1-2, pp.~285--293, 2017.

\bibitem{vock2019powders}
S.~Vock, B.~Kl{\"o}den, A.~Kirchner, T.~Wei{\ss}g{\"a}rber, and B.~Kieback,
  ``Powders for powder bed fusion: a review,'' {\em Progress in Additive
  Manufacturing}, vol.~4, pp.~383--397, 2019.

\bibitem{wu2019modeling}
Y.~Wu and G.~Chiu, ``Modeling height profile for drop-on-demand print of uv
  curable ink,'' in {\em Dynamic Systems and Control Conference}, vol.~59155,
  p.~V002T13A006, American Society of Mechanical Engineers, 2019.

\bibitem{carou2024inkjet}
P.~Carou-Senra, L.~Rodr{\'\i}guez-Pombo, A.~Awad, A.~W. Basit,
  C.~Alvarez-Lorenzo, and A.~Goyanes, ``Inkjet printing of pharmaceuticals,''
  {\em Advanced Materials}, vol.~36, no.~11, p.~2309164, 2024.

\bibitem{daly2015inkjet}
R.~Daly, T.~S. Harrington, G.~D. Martin, and I.~M. Hutchings, ``Inkjet printing
  for pharmaceutics--a review of research and manufacturing,'' {\em
  International journal of pharmaceutics}, vol.~494, no.~2, pp.~554--567, 2015.

\bibitem{goyanes2015personalized}
A.~Goyanes, J.~Wang, A.~Buanz, R.~Mart{\'\i}nez-Pacheco, R.~Telford,
  S.~Gaisford, and A.~W. Basit, ``3d printing of medicines: engineering novel
  oral devices with unique design and drug release characteristics,'' {\em
  Molecular pharmaceutics}, vol.~12, no.~11, pp.~4077--4084, 2015.

\bibitem{bernard2018biocompatibility}
M.~Bernard, E.~Jubeli, M.~D. Pungente, and N.~Yagoubi, ``Biocompatibility of
  polymer-based biomaterials and medical devices--regulations, in vitro
  screening and risk-management,'' {\em Biomaterials science}, vol.~6, no.~8,
  pp.~2025--2053, 2018.

\bibitem{wu2021improved}
Y.~Wu and G.~Chiu, ``An improved model of height profile for drop-on-demand
  print of ultraviolet curable ink,'' {\em ASME Letters in Dynamic Systems and
  Control}, vol.~1, no.~3, p.~031010, 2021.

\bibitem{chou2021inkjet}
W.-H. Chou, A.~Gamboa, and J.~O. Morales, ``Inkjet printing of small molecules,
  biologics, and nanoparticles,'' {\em International Journal of Pharmaceutics},
  vol.~600, p.~120462, 2021.

\bibitem{wang2023review}
S.~Wang, X.~Chen, X.~Han, X.~Hong, X.~Li, H.~Zhang, M.~Li, Z.~Wang, and
  A.~Zheng, ``A review of 3d printing technology in pharmaceutics: Technology
  and applications, now and future,'' {\em Pharmaceutics}, vol.~15, no.~2,
  p.~416, 2023.

\bibitem{borandeh2021polymeric}
S.~Borandeh, B.~van Bochove, A.~Teotia, and J.~Sepp{\"a}l{\"a}, ``Polymeric
  drug delivery systems by additive manufacturing,'' {\em Advanced drug
  delivery reviews}, vol.~173, pp.~349--373, 2021.

\bibitem{seoane2023case}
I.~Seoane-Viano, X.~Xu, J.~J. Ong, A.~Teyeb, S.~Gaisford,
  A.~Campos-{\'A}lvarez, A.~Stulz, C.~Marcuta, L.~Kraschew, W.~Mohr, {\em
  et~al.}, ``A case study on decentralized manufacturing of 3d printed
  medicines,'' {\em International Journal of Pharmaceutics: X}, vol.~5,
  p.~100184, 2023.

\bibitem{lupi2023blockchain}
F.~Lupi, M.~G. Cimino, T.~Berlec, F.~A. Galatolo, M.~Corn, N.~Ro{\v{z}}man,
  A.~Rossi, and M.~Lanzetta, ``Blockchain-based shared additive
  manufacturing,'' {\em Computers \& Industrial Engineering}, vol.~183,
  p.~109497, 2023.

\bibitem{melocchi2016hot}
A.~Melocchi, F.~Parietti, A.~Maroni, A.~Foppoli, A.~Gazzaniga, and L.~Zema,
  ``Hot-melt extruded filaments based on pharmaceutical grade polymers for 3d
  printing by fused deposition modeling,'' {\em International journal of
  pharmaceutics}, vol.~509, no.~1-2, pp.~255--263, 2016.

\bibitem{wang2021emerging}
J.~Wang, Y.~Zhang, N.~H. Aghda, A.~R. Pillai, R.~Thakkar, A.~Nokhodchi, and
  M.~Maniruzzaman, ``Emerging 3d printing technologies for drug delivery
  devices: Current status and future perspective,'' {\em Advanced drug delivery
  reviews}, vol.~174, pp.~294--316, 2021.

\bibitem{govender2020independent}
R.~Govender, S.~Abrahms{\'e}n-Alami, A.~Larsson, A.~Borde, A.~Liljeblad, and
  S.~Folestad, ``Independent tailoring of dose and drug release via a
  modularized product design concept for mass customization,'' {\em
  Pharmaceutics}, vol.~12, no.~8, p.~771, 2020.

\bibitem{wu2021improved2}
Y.~Wu and G.~Chiu, ``An improved height difference based model of height
  profile for drop-on-demand 3d printing with uv curable ink,'' in {\em 2021
  American Control Conference (ACC)}, pp.~491--495, IEEE, 2021.

\bibitem{melocchi20153d}
A.~Melocchi, F.~Parietti, G.~Loreti, A.~Maroni, A.~Gazzaniga, and L.~Zema, ``3d
  printing by fused deposition modeling (fdm) of a swellable/erodible capsular
  device for oral pulsatile release of drugs,'' {\em Journal of Drug Delivery
  Science and Technology}, vol.~30, pp.~360--367, 2015.

\bibitem{trivedi2018additive}
M.~Trivedi, J.~Jee, S.~Silva, C.~Blomgren, V.~M. Pontinha, D.~L. Dixon,
  B.~Van~Tassel, M.~J. Bortner, C.~Williams, E.~Gilmer, {\em et~al.},
  ``Additive manufacturing of pharmaceuticals for precision medicine
  applications: A review of the promises and perils in implementation,'' {\em
  Additive Manufacturing}, vol.~23, pp.~319--328, 2018.

\bibitem{attaran2017additive}
M.~Attaran {\em et~al.}, ``Additive manufacturing: the most promising
  technology to alter the supply chain and logistics,'' {\em Journal of Service
  Science and Management}, vol.~10, no.~03, p.~189, 2017.

\bibitem{rafi2022regulatory}
K.~Rafi, A.~Zhonghong~Liu, M.~Di~Prima, P.~Bates, and M.~Seifi, ``Regulatory
  and standards development in medical additive manufacturing,'' {\em MRS
  Bulletin}, vol.~47, no.~1, pp.~98--105, 2022.

\bibitem{geraili2021design}
A.~Geraili, M.~Xing, and K.~Mequanint, ``Design and fabrication of
  drug-delivery systems toward adjustable release profiles for personalized
  treatment,'' {\em View}, vol.~2, no.~5, p.~20200126, 2021.

\bibitem{wu2022height}
Y.~Wu, {\em Height profile modeling and control of inkjet 3d printing}.
\newblock PhD thesis, Purdue University, 2022.

\bibitem{cailleaux2021fused}
S.~Cailleaux, N.~M. Sanchez-Ballester, Y.~A. Gueche, B.~Bataille, and
  I.~Soulairol, ``Fused deposition modeling (fdm), the new asset for the
  production of tailored medicines,'' {\em Journal of controlled release},
  vol.~330, pp.~821--841, 2021.

\bibitem{kapoor2020coating}
D.~Kapoor, R.~Maheshwari, K.~Verma, S.~Sharma, P.~Ghode, and R.~K. Tekade,
  ``Coating technologies in pharmaceutical product development,'' in {\em Drug
  delivery systems}, pp.~665--719, Elsevier, 2020.

\bibitem{wu2023error}
Y.~Wu and G.~Chiu, ``Error diffusion based feedforward height control for
  inkjet 3d printing,'' in {\em 2023 IEEE/ASME International Conference on
  Advanced Intelligent Mechatronics (AIM)}, pp.~125--131, IEEE, 2023.

\bibitem{boehm2014inkjet}
R.~D. Boehm, P.~R. Miller, J.~Daniels, S.~Stafslien, and R.~J. Narayan,
  ``Inkjet printing for pharmaceutical applications,'' {\em Materials Today},
  vol.~17, no.~5, pp.~247--252, 2014.

\end{thebibliography}

\end{document}